\documentclass[twocolumn,showpacs,preprintnumbers,amsmath,amssymb]{revtex4}

\usepackage{graphicx}
\usepackage{dcolumn}
\usepackage{bm}

\begin{document}


\title{Atomic Transport in Dense, Multi-Component Metallic Liquids}

\author{A.\ Meyer}
 \email{ameyer@ph.tum.de}
 \affiliation{Physik Department E\,13, Technische Universit\"at M\"unchen, 85747 
Garching, Germany}

\date{\today: Physical Review E (submitted) and cond-mat/0206364}

\begin{abstract}
Pd$_{43}$Ni$_{10}$\-Cu$_{27}$\-P$_{20}$
has been investigated in its equilibrium liquid state with incoherent, 
inelastic neutron scattering.
As compared to simple liquids, liquid PdNiCuP
is characterized by a dense packing with a packing fraction above 0.5.
The intermediate scattering function exhibits a fast relaxation process
that precedes structural relaxation. 
Structural relaxation obeys a time-temperature superposition
that extends over a temperature range of 540\,K. 
The mode-coupling theory of the liquid to glass transition (MCT)
gives a consistent description of the dynamics which governs
the mass transport in liquid PdNiCuP alloys.
MCT scaling laws extrapolate to a critical temperature 
$T_c$ at about 20\,\% below the liquidus temperature.
Diffusivities derived from the mean relaxation times 
compare well with Co diffusivities from recent tracer diffusion measurements
and diffsuivities calculated from viscosity via the Stokes-Einstein relation.
In contrast to simple metallic liquids, the atomic transport in dense,
liquid Pd$_{43}$Ni$_{10}$\-Cu$_{27}$\-P$_{20}$
is characterized by a drastical slowing down of dynamics on cooling,
a $q^{-2}$ dependence of the mean relaxation times at intermediate $q$ 
and a vanishing isotope effect as a result of a highly collective transport 
mechanism.
At temperatures as high as $2\!\times\!T_c$
diffusion in liquid Pd$_{43}$Ni$_{10}$\-Cu$_{27}$\-P$_{20}$
is as fast as in simple liquids at the melting point.
However, the difference in the underlying atomic transport mechanism 
indicates that the diffusion mechanism in liquids 
is not controlled by the value of the diffusivity but rather by that of the
packing fraction.

\end{abstract}

\pacs{61.25.Mv,61.20.Lc,61.12.-q}

\maketitle

\section{\label{intro}Introduction}

We investigate microscopic dynamics in liquid PdNiCuP melts with
inelastic neutron scattering. Our results show that PdNiCuP can in 
good approximation be regarded as multicomponent hard-sphere like system.
As compared to liquid alkali-metals the packing in liquid PdNiCuP is
much more dense.
As a consequence, dynamics in liquid PdNiCuP can not be described
by concepts developed for simple liquids. Instead,
atomic transport in PdNiCuP melts is in excellent accordance
with concepts developed in the context of glass formation. 

Because of the short range nature of the interatomic potential  
with a strong repulsive core, 
the potential in simple metals can in good approximation be condensed to an 
effective hard sphere radius $R$. Therefore, the packing fraction
$\varphi\!=\!\frac{4}{3}\pi\,n\, R^3$ ($n$ is the number density in unit volume)
is an important parameter for the discussion of the transport mechanism \cite{PrAP73}.
Due to the low compressibility of liquid metals the hard sphere radius
is fairly temperature independent \cite{Pas88}.
Liquid alkali metals are a paradigm of hard-sphere like fluids and their
microscopic dynamics have been investigated in great detail \cite{BaZo94,TyHa84}.
At low packing fraction, e.g.\ at temperatures well above the 
melting temperature,
atomic transport is dominated by binary collisions.
This reflects itself in a Gaussian line shape of the quasielastic 
neutron scattering signal in the free particle limit towards large 
wavenumbers $q$.
With increasing packing fraction, i.e.\ by approaching the melting temperature $T_m$,
fluid dynamics play a more important role
and towards small but finite $q$ the quasielastic line is well approximated
by a Lorentzian function \cite{MoGl86,BaZo94}.
Alkali melts exhibit a packing fraction of about 0.4 at the melting point. 

Colloidal suspensions \cite{MeUn93}
and molecular dynamics simulated binary metallic melts
\cite{KoAn94,Tei96}, that have a significantly higher packing fraction above
0.5, exhibit microscopic dynamics that is not governed by binary collisions and is 
typical for glass forming, 
molecular liquids \cite{WoAn76,CuLH97,Goe99}. Their dynamics has been described by 
predictions of the mode coupling theory 
(MCT) of the liquid to glass transition \cite{GoSj92}:
Atomic transport slows down by orders of magnitude on a small change in temperature
or pressure.
The slowing down of the dynamics goes along with a spread of the quasielastic line as 
compared to the Lorentzian found in simple liquids 
corresponding to a stretching in time of correlation functions over a wider time 
range than expected for an exponential decay.

Mode coupling theory gives a microscopic explanation for this behaviour in terms of a
non linear coupling of density fluctuations caused by feedback effects in the dense
liquid. Atomic transport is envisioned as a highly collective process.
At a critical packing fraction $\varphi_c$, or a critical temperature $T_c$ respectively, 
MCT predicts a change in the transport mechanism from liquid-like flow 
to glass like activated hopping processes.
MCT calculations for a hard sphere system exhibit a critical packing fraction
of 0.525 \cite{FuHL92}. In a colloidal suspension $\varphi_c \simeq 0.58$ has been
found \cite{MeUn93}.
Whereas in alkali melts the packing fractions are well below 0.5, 
new multicomponent alloys on Zr- \cite{PeJo93} and Pd-basis \cite{InNK97,LuWG99,NiIn02}
at quasieutectic compositions exhibit packing fractions at the their liquidus temperature 
$T_{liq}$ of some 0.52 \cite{OhCR97,LuGF02}.
These alloys known for their bulk glass forming ability, exhibit viscosities
at their melting temperatures that are 2-3 orders of magnitude larger 
than in simple metals and most alloys \cite{MaWB99,HaBD02}.

In previous neutron scattering experiments fast dynamics in 
viscous Zr$_{46.8}$Ti$_{8.2}$Cu$_{7.5}$Ni$_{10}$Be$_{27.5}$ have been investigated:
Data analysis in the framework of MCT is in full agreement with the predicted scaling 
functions and gives a $T_c$ some 20\,\% below $T_{liq}$ \cite{MeWP98}.
In an experiment on viscous Pd$_{40}$Ni$_{10}$\-Cu$_{30}$\-P$_{20}$ (Pd40) the focus was 
on slow dynamics \cite{MeBS99}:
The long--time decay of correlations exhibits the common 
features of glass forming liquids, i.e.\ 
structural relaxation obeys a stretching in time and a universal 
time-temperature superposition.

Here, inelastic neutron scattering results on
liquid Pd$_{43}$Ni$_{10}$\-Cu$_{27}$\-P$_{20}$ (Pd43) are reported.
The small change in composition compared to the
Pd$_{40}$Ni$_{10}$\-Cu$_{30}$\-P$_{20}$
results in an improved stability with respect to crystallization
\cite{LuWG99} ---
a cooling rate as low as 0.09\,K/s 
is sufficient to avoid crystallization and to form bulk metallic 
glass \cite{ScJB00}.
This allows measurements of transport coefficients at temperatures around the
mode coupling $T_c$ with experimental techniques that only require heat treatment
for some minutes, e.g.\ rheometry \cite{MaWB99} or 
radio tracer diffusion measurements \cite{Zoe02}.

Compared to previous neutron scattering measurements \cite{MeWP98,MeBS99}
a different experimental setup has been used 
that covers an extended range of momentum transfers $q$.
This allows a detailed investigation of the fast MCT $\beta$ relaxation and the 
correlation between structural relaxation 
and long range atomic transport.
In addition, the temperature range has been extended by several 100\,K  
up to temperatures at which the diffusion of the atoms is of the order of
that in simple monoatomic metals.
The results are set in context to macroscopic transport coefficients,
the dynamics in simple liquids
and the predictions by the mode coupling theory of the
liquid to glass transition.

\section{\label{exp}Experimental}

Pd$_{43}$Ni$_{10}$\-Cu$_{27}$\-P$_{20}$ was prepared
from a mixture of pure elements by
induction melting in a silica tube.
The melt was subject to a B$_2$O$_3$ flux treatment in order 
to remove oxide impurities.
Differential scanning calorimetry with a heating rate of 40\,K/min
resulted in a $T_g$ at 578\,K and a $T_{liq}$ at 863\,K
in accordance with literature values \cite{LuWG99,ScJB00}.
For the neutron time-of-flight experiment a thin-walled Al$_2$O$_3$ container 
has been used giving a hollow cylinder sample geometry of 22\,mm in diameter
and a thickness of 1\,mm. 
During the measurement the liquid was covered by a thin layer of 
$^{11}$B$_2$O$_3$ flux material.
For the chosen geometry and neutron wavelength the sample scatters
less than 2\%. 
Therefore, multiple scattering, which would alter the data
especially towards low $q$ \cite{Wut00}, only has a negligible effect.

Microscopic dynamics in liquid PdNiCuP
has been investigated
on the neutron time-of-flight spectrometer IN\,6 at the 
Institut Laue-Langevin in Grenoble. 
An incident neutron wavelength of 
$\lambda\!=\!5.1\,\mbox{\AA}^{-1}$ 
yielded an accessible wavenumber range at zero energy transfer
of $q\!=\!0.75-1.95\,\mbox{\AA}^{-1}$ at an energy resolution of 
$92\,\mu\mbox{eV}$ (FWHM).
Regarding the scattering cross sections of the individual elements
PdNiCuP is a 90\,\% coherent scatterer. 
However, with the first structure factor maximum at 
$q_{0}\!\simeq\!2.9$\,\AA$^{-1}$  
our spectra are dominated by incoherent scattering, 
that is dominated by the contributions from Ni with 
$\simeq$\,73\,\% and Cu with $\simeq$\,20\,\%.

\begin{figure}[t]
\includegraphics[scale=0.75]{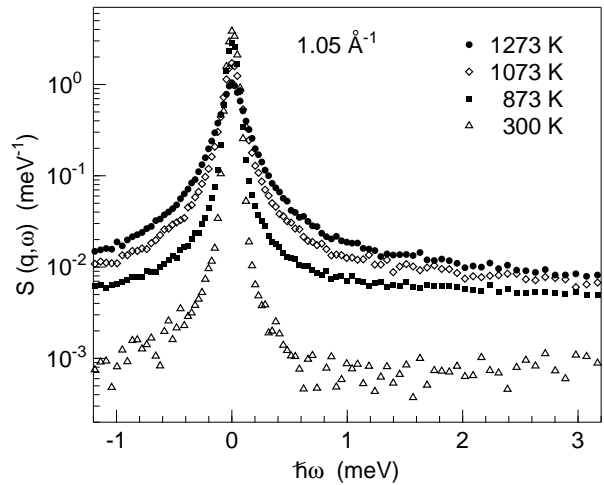}
\caption{\label{sqw} 
Scattering law $S(q,\omega)$ (logarithmic scale) of liquid 
Pd$_{43}$Ni$_{10}$\-Cu$_{27}$\-P$_{20}$ 
obtained on the neutron time-of-flight spectrometer IN\,6. 
The signal is dominated by the incoherent scattering from Ni
and Cu.
The data at 300\,K represent the instrumental energy resolution.
Measurements in the liquid ($T_{liq}\!=\!863\,$K) 
reveals a broad quasielastic signal.
}
\end{figure}

The alloy was measured at room temperature to obtain the
instrumental energy resolution function. 
Data were collected in the liquid between 833\,K and 913\,K 
in steps of 40\,K and
between 973\,K and 1373\,K in steps of 100\,K with a duration
between 2 and 5 hours each.
At each temperature empty cell runs were performed.
The data at 833\,K, 30\,K below the liquidus, do not show
signs of crystallization, which is in accordance with
the time-temperature transformation of the undercooled liquid \cite{ScJB00}.
In order to obtain the scattering law $S(q,\omega)$ (Fig.~\ref{sqw}), raw data were
normalized to a Vanadium standard, corrected for self-absorption
and container scattering, interpolated to constant $q$, and
symmetrized with respect to energy with the detailed balance factor.
Fourier deconvolution of $S(q,\omega)$ and normalization to
1 for $t=0$ gives the self correlation function
$\Phi(q,t)$.


\section{\label{mct}Mode-Coupling Theory Asymptotics}

The mode-coupling theory of the liquid to glass transition (MCT)
\cite{GoSj92,Goe99} is developed in the
well-defined frame of molecular hydrodynamics \cite{BoYi80}.
Starting with the particle density, $\vec{\rho_r}(t)= \sum_{i} 
\delta(\vec{r} - \vec{R}_i(t))$, i.e.\ the positions, $\vec{R}_i$,
of each particle, $i$, at time~$t$,
the Zwanzig-Mori formalism provides an exact equation of motion for
the density correlation function, 
$\Phi_q(t) = 
\langle\rho_q (t)^* \rho_q\rangle \,/\,\langle | \rho_q | ^2 \rangle$:
\begin{equation}\label{fullmct}
\ddot{\Phi}_q (t) + \Omega_q^2\, \Phi_q (t) + 
  \int_{0}^{t} M_q (t-t') \,\dot{\Phi}_q (t') \mbox{d}t' = 0
\end{equation}
where $\Omega_q$ represents a phonon frequency at wavenumber $q$ 
and $M_q(t)$ a correlation function of force fluctuations which in turn 
is a functional of density correlations.
The static part of the fluctuating force is a linear combination 
of density fluctuation pairs and depends therefore only on the
interaction potential.

Since MCT does not aim to describe the detailed microscopic phonon 
distribution, but rather to give a universal picture
of relaxational dynamics at longer times, the integral kernel 
is split into 
\begin{equation}\label{split}
M_q(t) = \nu_q\delta(t) + \Omega_q^2 m_q(t),
\end{equation}
where $\nu_q$ models the damping by ``fast'' modes and $m_q(t)$
accounts for memory effects through the coupling of 
``slow'' modes.
The basic idea of the mode-coupling theory of the liquid to glass transition
is to consider as ``slow'' all products of density fluctuations.
By derivation, $m_q(t)$ contains no terms linear in $\Phi_q(t)$. 
Therefore, in lowest order, it is a quadratic functional
\begin{equation}\label{kernel}
m_q (t) = \sum_{q_1+q_2=q} V_q(q_1,q_2) \,\Phi_{q_1}(t) \,\Phi_{q_2}(t). 
\end{equation}
In this approximation, the coupling coefficients $V_q$ are specified in terms 
of the static structure of the liquid.

Eqs.~1, 2 and 3 lead to the following scenario:
a fast $\beta$-relaxation process, which can be visualized as 
a rattling of the atoms in the cages formed by their neighbouring atoms,
prepares structural $\alpha$ relaxation, responsible for viscous flow.
At an ideal glass transition temperature $T_c$ the transport 
mechanism crosses over from glass-like activated hopping processes to liquid-like 
collective motion. 
In other words, at $T_c$, the cages are no longer stable on the 
time scale of a diffusive jump.

A common feature of structural $\alpha$ relaxation in glass-forming 
liquids is a stretching in correlation functions over a wider time range 
than expected for exponential relaxation \cite{WoAn76,CuLH97}.
Experimental data in the $\alpha$ relaxation regime can usually be well 
described by a stretched exponential function
\begin{equation}\label{kww}
F(q,t) = f_q^c\,\exp{[-(t/\tau_q)}^{\beta_q}]
\end{equation}
with an exponent $\beta_q\!<\!1$. $\tau_q$ is the relaxation time
and $f_q^c\!<\!1$ accounts for the initial decay of
correlations due to phonons and the fast relaxation process.
For temperatures above $T_c$ mode coupling theory predicts 
a universal time-temperature superpostion of structural relaxation
resulting in a temperature independent stretching exponent $\beta_q$.
Many glass-forming liquids in contrast exhibit a temperature dependence of
$\beta$ -- in some $\beta$ is decreasing, in others increasing on 
temperature increase.

Solutions of the mode coupling equations for a hard sphere system
confirm, that the MCT $\alpha$ relaxation is well described by 
a stretched exponential function and  
Eq.\ (\ref{kww}) becomes a special, long time solution of 
mode coupling theory \cite{FuHL92}. 
For the mean relaxation times
\begin{equation}\label{mtau}
  \left\langle\tau_q\right\rangle = \int\limits_0^\infty\! {\rm d}t\,
    F(q,t) / f_q^c = \tau_q \, 
   \beta^{-1}\Gamma ({\beta^{-1}})
\end{equation}
mode coupling theory predicts that $\left\langle\tau_q(T)\right\rangle$ 
is inversely proportional to the diffusivity $D(T)$ and obeys an asymptotic scaling law
for temperatures above but close to $T_c$:
\begin{equation}\label{tauscal}
\langle\tau_q\rangle \propto 1 / D \propto [(T\!-\!T_c)/T_c]^{-\gamma}.
\end{equation}

Asymptotic expansions of Eq.\ (\ref{fullmct},\ref{kernel})
show that in the
intermediate $\beta$-relaxation regime around a crossover time,
$t_{\sigma}$, and for temperatures close to $T_c$,
the asymptotic form of the correlation function is 
independent of the detailed structure of the coupling 
coefficients and exhibits a universal factorization property:
\begin{equation}\label{betascal}
\Phi(q,t)= f_q^c + h_q\,g_{\rm \lambda}(t/t_{\sigma}),
\end{equation}
where $f_q$ represents the Debye-Waller factor and $h_q$
an amplitude. The scaling function 
$g_{\rm \lambda}(\tilde{t})$ is defined by just one shape 
parameter $\lambda$.
Close to $T_c$ mode coupling theory predicts 
the temperature dependence of $t_{\sigma}$ and $h_q$ 
with the asymptotic scaling functions:
\begin{equation}\label{tchq}
   t_\sigma \propto (T-T_c)^{-1/2a} \quad{\rm and}\quad
   h_q      \propto (T-T_c)^{1/2}.
\end{equation}
There are tables providing $a$, $\gamma$ and $g_{\lambda}$
as a function of $\lambda$~\cite{Goe90}.

\section{\label{res}Results}

Figure \ref{sqw} shows the scattering law $S(q,\omega)$ of
liquid PdNiCuP at $q\!=\!1.05$\,\AA$^{-1}$.
The signal is dominated by the incoherent scattering from Ni
and Cu. $S(q,\omega)$ displays 
a quasielastic line with an increasing width on temperature increase
and with wings extending up to several meV.
Above some 3\,meV the quasielastic signal merges into a constant as expected for 
an ideal Debye solid and found in liquid GeO$_2$ \cite{MeSN01}.  
The quasielastic signal of $S(q,\omega)$ is fairly well separated from
vibrations.

\subsection{\label{vib}Vibrations}

There is no general understanding of the influence of phonons on the 
atomic transport in liquids. 
Whereas in monoatomic alkali melts one finds low-lying collective
phonon modes that mediate mass transport \cite{WaSj82,MoGG87,BaZo94}, 
the mode coupling theory of the liquid to glass transition does not consider
the detailed phonon dynamics for its description of atomic transport 
in viscous liquids (Eq.\ \ref{split}).
In particular, dynamics in hard sphere-like colloidal suspensions, 
that do not exhibit vibrations, are in excellent agreement with 
MCT predictions \cite{MeUn93}. 

\begin{figure}[t]
\includegraphics[scale=0.75]{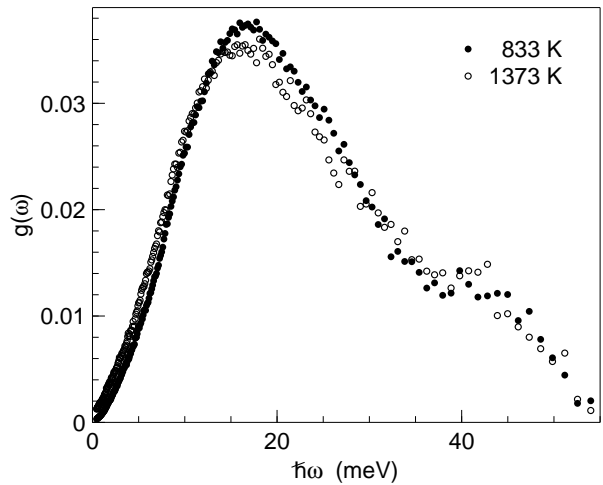}
\caption{\label{gw} 
Density of states $g(\omega)$ representing mainly the Ni and Cu vibrations in liquid 
Pd$_{43}$Ni$_{10}$\-Cu$_{27}$\-P$_{20}$. 
With a maximum at some 17\,meV (corresponding to some 0.04\,ps)
vibrations are fairly well separated
from the fast relaxational dynamics around 1\,ps.
The small expansion coefficient reflects itself in a weak temperature
dependence of $g(\omega)$.}
\end{figure}

In most molecular liquids broad phonon distributions overlap with 
the quasielastic signal.
The phonon density of states exhibits a first maximum usually 
between 4 and 8\,meV \cite{WuPC95,Tol01}.
This aggravates the theoretical description of the relaxational
dynamics.
Figure \ref{gw} displays the vibrational density of states $g(\omega)$
representing mainly the incoherent contributions of the Ni and Cu atoms
of liquid PdNiCuP. 
$g(\omega)$ has been derived from the scattering law $S(q,\omega)$
using a procedure that assumes pure incoherent scattering
and uses the multiphonon correction as described
in \cite{WuKB93}.

$g(\omega)$ displays a weak temperature dependence indicating that
anharmonic contributions to the interatomic potential are minor.
This is in accordance with an expansion coefficient of only
$4 \times 10^{-5}$\,K$^{-1}$ \cite{LuGF02}, that is an order of magnitude 
smaller than that of most molecular glass forming liquids.
With a maximum in $g(\omega)$ at some 17\,meV,  
vibrations are fairly well separated
from the quasielastic signal that extends up to several meV.
In addition, PdNiCuP does not exhibit a ``boson peak'' --
a maximum in $S(q,\omega)$ that is found in other glasses 
usually at a few meV -- in the $q$ range investigated.
This allows a detailed investigation of the atomic transport 
in liquid PdNiCuP.


\subsection{\label{alpha} Structural Relaxation}

\begin{figure}[b]
\includegraphics[scale=0.75]{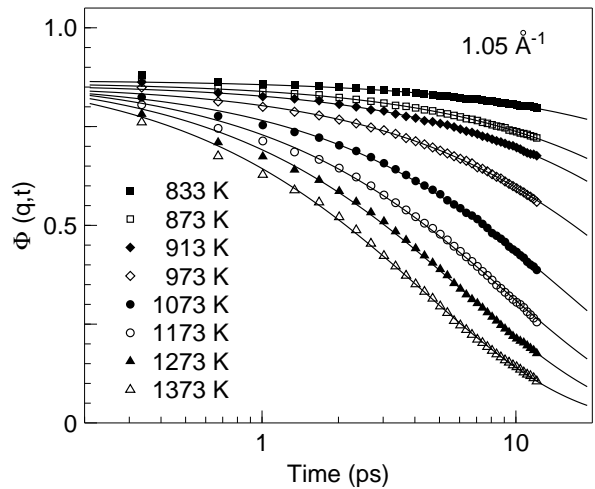}
\caption{\label{sqt} Normalized density correlation function $\Phi(q,t)$
of liquid PdNiCuP at 1.05\,\AA$^{-1}$ as obtained
by Fourier deconvolution of measured $S(q,\omega)$.
Structural relaxation causes the final decay of $\Phi(q,t) \to 0$.
The lines are fits a stretched exponential function (Eq.\ \ref{kww}).}
\end{figure}

The quasielastic signal is best analyzed in the time domain with a removal of
the instrumental resolution function.
The density correlation function $\Phi(q,t)$ has been obtained by 
Fourier transformation of measured $S(q,\omega)$,
division of the instrumental resolution function, 
and normalization with the value at $t=0$.
Between 0 and $\simeq\!1$\,ps phonons and a fast process
lead to a decrease in $\Phi(q,t)$ from 1 towards a plateau.
Figure \ref{sqt} displays the long-time decay of 
$\Phi(q,t)$ at $q\!=\!1.05$\,\AA$^{-1}$ from this plateau 
towards zero in a semilogarithmic representation.

\begin{figure}[t]
\includegraphics[scale=0.75]{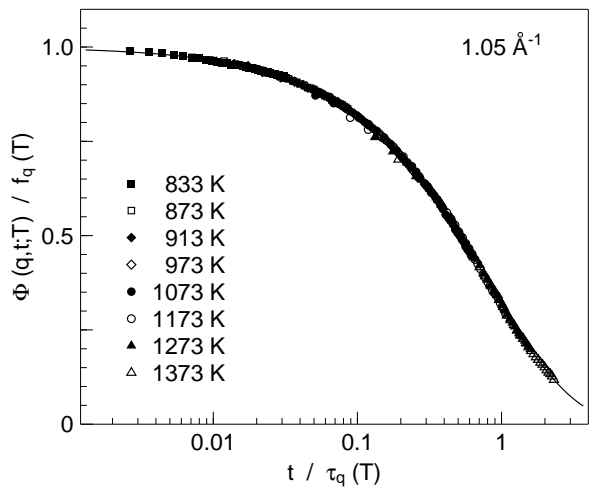}
\caption{\label{tts} Rescaling of the density correlation function in the $\alpha$ 
relaxation regime (for $t > 1$\,ps) 
using results from fits with a stretched exponential function: a time-temperature superposition
of structural relaxation holds from $T_{liq} - 30$\,K to $T_{liq} + 510$\,K.
The line is a fit with Eq.\ \ref{kww} resulting in a stretching exponent 
$\beta\!=\!0.75$.}
\end{figure}

\begin{figure}[h]
\includegraphics[scale=0.75]{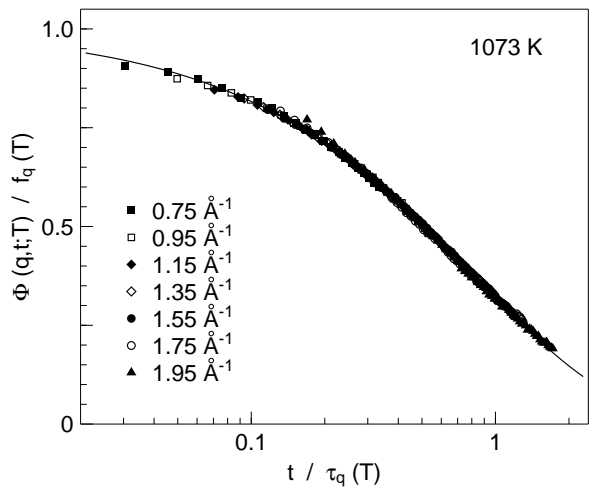}
\caption{\label{tqs} Rescaling of the density correlation function $\Phi(q,t)$
at 1073\,K. Within the accessible $q$ range the stretching of the
structural relaxation is fairly $q$ independent. 
The line represents a stretched exponential function (Eq.\ \ref{kww}) 
with a stretching exponent 
$\beta\!=\!0.75$.}
\end{figure}

In contrast to simple metallic liquids, liquid PdNiCuP
exhibits a structural relaxation that shows stretching in time.
The lines are fits with the stretched exponential function (Eq.\ \ref{kww}).
In a first fitting process the stretching exponent $\beta_q(T)$ was treated as a 
parameter: $\beta_q(T)$ varies weakly around a mean
$\beta\!=\!0.75$. For the further data analysis a $\beta\!=\!0.75$ was used.

Structural relaxation described by mode coupling theory displays stretching
and a time-temperature superposition.
Using results from the fitting procedure with Eq.\ \ref{kww} and a stretching
exponent $\beta\!=\!0.75$ master curves $\Phi(q,(t/\tau_q))/f_q$ have been constructed 
for the data in the structural relaxation regime above some 1\,ps.
Figure \ref{tts} shows rescaled $\Phi(q,(t/\tau_q))/f_q$ at $q\!=\!1.05\,$\AA$^{-1}$.
Over the entire temperature range, that spans 540\,K,
the $\Phi(q,(t/\tau_q))/f_q$ fall on a master curve: 
a time-temperature superposition of structural relaxation holds.
We note, that the validity of the time-temperature superposition
up to temperatures at which the transport coefficients approach
that of simple liquids (\ref{diffsec}) is in marked contrast to
the idea, that viscous liquids exhibit a transition to a liquid
with the stretching exponent $\beta$ approaching 1 on temperature 
increase.

Stretching of self correlation functions is generally found to be more 
pronounced in fragile glass-forming liquids, 
characterized by a curved temperature dependence of viscosity 
in an Arrhenius plot \cite{BoNA93}; 
e.g.\ the van der Waals liquid orthoterphenyl 
with a $\beta\!\simeq\!0.5$ \cite{Tol01}.
In an intermediate system like hydrogen-bond forming glycerol 
a $\beta\!\simeq\!0.6$ has been reported \cite{WuCR96}.
The stretching exponent in liquid PdNiCuP
metals compares well to the value $\beta\!\simeq\!0.75$ found
in covalent-network forming sodium disilicate melts \cite{MeSD02}
and indicates that in this context the PdNiCuP alloy might classify as a fairly strong glass 
forming liquid. 

Figure \ref{tqs} shows rescaled $\Phi(q,(t/\tau_q))/f_q$ at 1073\,K 
for $q$ values in the range between 0.75\,\AA$^{-1}$ and 1.95\,\AA$^{-1}$.
Structural relaxation in liquid PdNiCuP
can be described with a stretched exponential function and a $q$ and
temperature independent stretching exponent $\beta = 0.75 \pm 0.02$.
$\beta_q$ shows a small but systematic decrease with increasing $q$ in agreement 
with MCT calculations for a hard-sphere system \cite{FuHL92}. 
However, a small variation of $\beta_q$ in the fitting procedure has 
no significant effect on the resulting mean relaxation times (Eq. \ref{mtau}).

\subsection{Diffusion}\label{diffsec}

\begin{figure}[b]
\includegraphics[scale=0.75]{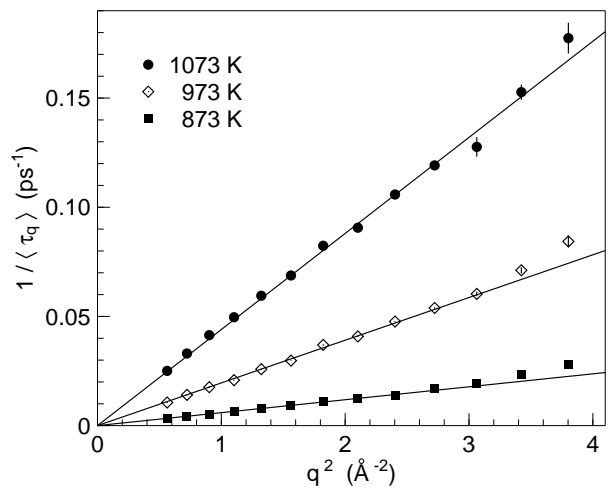}
\caption{\label{tauq}  Mean relaxation times $\langle\tau_q\rangle$ 
of tagged particle motion in liquid Pd$_{43}$Ni$_{10}$\-Cu$_{27}$\-P$_{20}$.
$1\,/\,\langle\tau_q\rangle$ shows a $q^2$ dependence as expected for
long range atomic transport for $q \to 0$. The slope corresponds
to the self diffusion coefficient $D(T)$.}
\end{figure}

In the hydrodynamic limit for $q \rightarrow 0$
one expects that the mean relaxation times $\langle\tau_q\rangle$ are indirect proportional
to the square of the momentum transfer $q$ \cite{BoYi80}. 
In liquid PdNiCuP the $1\,/\,q^2$ dependence holds even up to 1.9\,\AA$^{-1}$  
(Fig.\ \ref{tauq}) while
the first structure factor maximum is at $q_0\!\simeq\!2.9$\,\AA$^{-1}$.
In liquid alkali-metals one observes a systematic deviation from a $q^2$
dependence of the quasielastic line width 
already at $q$ values that correspond to 1\,/\,10 of their structure factor 
maximum $q_0$.
This deviation is explained by low-lying phonon modes that mediate atomic transport 
\cite{MoGG87,BaZo94}.
Mean relaxation times for self motion in the MCT hard sphere system, 
in contrast, vary rather well proportionally to $1\,/\,q^2$ for 
$q$ values extending even above the first structure factor maximum \cite{FuHL92}.
We note that mixing of hard spheres with different radii should enforce this behaviour even more.  
It appears that the validity of $1\,/\,\langle\tau_q\rangle \propto q^2$
also for intermediate $q$ values is a signature of the atomic transport mechanism
in dense liquids.

\begin{figure}[t]
\includegraphics[scale=0.75]{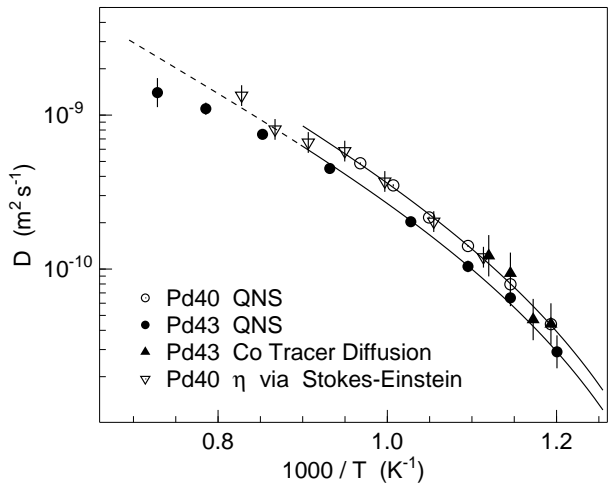}
\caption{\label{diff} 
Self diffusion coefficient $D$ of Ni and Cu in liquid Pd$_{43}$Ni$_{10}$\-Cu$_{27}$\-P$_{20}$
(closed circles) and  Pd$_{40}$Ni$_{10}$\-Cu$_{30}$\-P$_{20}$ (open circles from Ref.\ \cite{MeBS99}) 
derived from the mean relaxation times.
The lines are fits with MCT $\tau$ scaling (Eq.\ \ref{tauscal}).
The data are well described with $T_c=700\pm30$ and $\gamma=2.7\pm0.2$.\\
For comparison $^{57}$Co tracer diffusion in Pd$_{43}$Ni$_{10}$\-Cu$_{27}$\-P$_{20}$ 
(Ref.\ \cite{ZoRF02}) and diffusivities calculated from viscosity $\eta$ of 
liquid Pd$_{40}$Ni$_{10}$\-Cu$_{30}$\-P$_{20}$
(Ref.\ \cite{HaBD02}) via the Stokes-Einstein relation 
(Eq.\ \ref{stoke}) are shown.}
\end{figure}

The $1\,/\,q^2$ dependence of the mean relaxation times, 
also demonstrates that structural relaxation leads to
long range atomic transport with a diffusivity \cite{BoYi80}
\begin{equation}
D= (\langle\tau_q\rangle q^2)^{-1}\, .
\end{equation}
Figure \ref{diff} displays the diffusivities $D$ 
in liquid Pd$_{43}$Ni$_{10}$\-Cu$_{27}$\-P$_{20}$ as a function of $1\,/\,T$.
Values range from $3\pm1\!\times\!10^{-11}$\,m$^2$s$^{-1}$
at 833\,K to $1.4\pm0.3\!\times\!10^{-9}$\,m$^2$s$^{-1}$ at 1373\,K.
At the liquidus diffusion is about 2 orders of magnitude slower
as compared to simple metallic liquids and most alloys.
Diffusivities in liquid Pd40
show a similar temperature dependence but are larger by some 20\,\%.
The smaller mobility of the atoms in Pd43 appears to come along with 
the slightly better glass forming ability of the
Pd43 alloy.

The lines in Figure \ref{diff} represent fits with 
the MCT $\tau$ scaling law (Eq.\ \ref{tauscal}) to the diffusivities
$D(T)$.
Although Eq.\ \ref{tauscal} is only valid close to $T_c$,
Eq.\ \ref{tauscal} allows a rough estimate of the crossover 
temperature $T_c$ and the exponent $\gamma$.
A fit to $D(T)$ of liquid Pd43
at temperatures up to 1073\,K yields a $T_c\!=\!700\!\pm\!30$\,K and a 
$\gamma\!=\!2.7\!\pm\!0.2$.
The exponent $\gamma$ compares well to the $\gamma\!=\!2.62$ 
found in the MCT hard sphere system \cite{FuHL92}.
The temperature dependence of the diffusivity in liquid PdNiCuP alloys
is different from the $D \propto T^n$ behaviour (with $n\simeq2$) 
expected for uncorrelated binary collisions of hard spheres
\cite{PrAP73,KiYo87} and found with inelastic neutron scattering in expanded
liquid alkali-metals at high temperature \cite{WiPH93}.

Convection effects are a severe problem in macroscopic diffusion
measurements in ordinary liquids under gravity conditions 
with respect to the absolute value of the self diffusivity $D$ and its
temperature dependence.
A $\mu$g experiment on liquid Sn revealed that well above the melting point
convection even dominates the mass transport \cite{FrKW87}.
The $\mu$g data in liquid Sn obey $D \propto T^2$ for temperatures between 
$T_m$ and $2 \times T_m$. Liquid Sn exhibits a packing fraction
of $\simeq$\,0.4 as in liquid alkali-metals.
Inelastic neutron scattering data are not affected by convection 
because it probes dynamics on significantly shorter times.

Diffusivities from recent $^{57}$Co tracer diffusion measurements 
in liquid Pd43 \cite{Zoe02} 
are in excellent agreement
with the diffusivities obtained from inelastic neutron scattering (Fig.\ \ref{diff}).
This indicates that the mobility of Co is very similar to the Ni and Cu mobility 
observed in the incoherent neutron scattering signal and that
convection effects play no significant role in the tracer diffusion experiment.
The latter is in accordance with a viscosity $\eta$ that is two orders of magnitude 
larger as compared to simple metallic melts \cite{HaBD02}.

In simple liquids shear viscosity $\eta$ and the self diffusivity $D$ 
generally obey the Stokes-Einstein relation \cite{BaZo94,TyHa84,KiYo87}:
\begin{equation}\label{stoke}
D = k_b T / (6\pi a \eta),
\end{equation}
with reasonable values for the hydrodynamic radius $a$.  
Eq.\ \ref{stoke} even holds in most molecular liquids \cite{ChSi97}.
Figure \ref{diff} shows the diffusivity $D_{\eta}$ calculated from the
viscosity data of liquid Pd40 \cite{HaBD02}
via Eq.\ \ref{stoke} using $a=1.15$\,\AA\ that represents the Ni hard sphere 
radius  (Cu: 1.17\,\AA) from Ref.\ \cite{Pas88}. 
The Stokes-Einstein relation holds with $D_{\eta} = D$. 
We note, that $a$ is similar to the mean next nearest
neighbor distance displayed in the static structure factor $d = 2\pi / q_0 \simeq 1.1$\,\AA. 
In multicomponent liquids, that exhibit a dense packing 
of hard spheres with comparable hard sphere radii 
one expects a similar mobility of the different components.
In PdNiCuP the hard sphere radii of the different atoms have values within 20\,\%.
Because viscous flow represents the dynamics of all components,
the validity of Eq.\ \ref{stoke} in liquid PdNiCuP 
demonstrates that above $T_{liq}$ the mobility of the large and numerous
Pd atoms is quite similar to the Ni and Cu atoms \cite{V1}.

In colloidal suspensions the mean relaxation times $\langle\tau\rangle(\varphi)$
and the inverse of the diffusivity $1/D(\varphi)$ show the same dependence
as a function of the packing fraction.  
The slope for $\langle\tau(\varphi)\rangle$ and $1/D(\varphi)$
results in a $\gamma=2.7$ \cite{MaMM98}.
In molecular dynamics simulations on viscous NiP \cite{KoAn94},
in contrast, the asymptotic MCT $\tau$ scaling prediction (Eq.\ \ref{tauscal}) 
is violated: $\langle\tau(T)\rangle$ and $1/D(T)$ do not exhibit the same
temperature dependence.
Figures \ref{diff} and \ref{transall} demonstrate that in liquid PdNiCuP 
Eq.\ \ref{tauscal} holds quite well.


\subsection{Fast \boldmath$\beta$ relaxation}

\begin{figure}[b]
\includegraphics[scale=0.75]{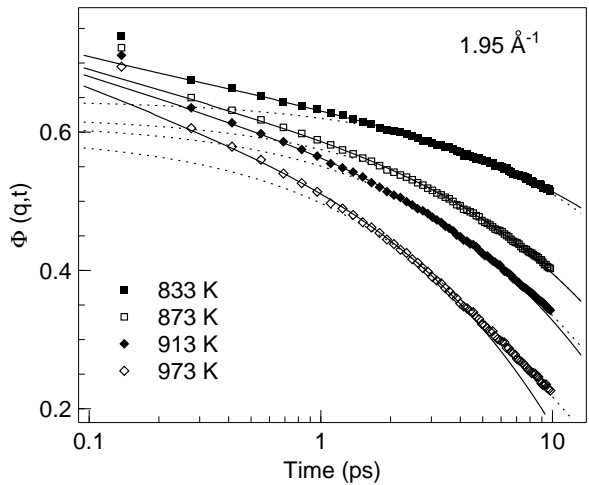}
\caption{\label{beta} Normalized time correlation function $\Phi(q,t)$ of liquid
Pd$_{43}$Ni$_{10}$\-Cu$_{27}$\-P$_{20}$ in the $\beta$ relaxation regime around 
one picosecond. The solid lines represent fits with the MCT $\beta$ scaling function
(Eq.\ \ref{betascal}) giving a temperature and $q$ independent shape parameter $\lambda = 0.78\!\pm\!0.04$.
Above $\simeq\,1$\,ps data are well described by a stretched exponential function 
(Eq.\ \ref{kww}, dashed lines).}
\end{figure}

Within the mode coupling theory of the liquid to
glass transition, structural relaxation and therefore long range mass transport
is preceded by a fast $\beta$ relaxation process whose time scale 
is typically in the order of a picosecond.
For $q$ values well below the structure factor maximum 
and incoherent scattering the amplitude of the $\beta$ relaxation is increasing with 
increasing $q$. 
Figure \ref{beta} displays the density correlation function of
liquid PdNiCuP at $q\!=\!1.95\,\AA^{-1}$.
Between 0 and $\simeq\!0.2$\,ps phonons lead to a rapid 
decay of atomic correlations (represented by the first data point in $\Phi(q,t)$ 
in Fig.\ \ref{beta}), so that $\Phi(q,t)$ decreases from
1 towards a plateau. 
On approaching this plateau a fast relaxation is clearly seen 
in $\Phi(q,t)$ through an additional intensity below some 1\,ps.
The lines are fits with the MCT $\beta$ scaling law (Eq.\ \ref{betascal}).
For $q<1.4$\,\AA$^{-1}$ the limited dynamic range of the instrument prevents
data analysis in the fast $\beta$ relaxation regime. 
Above 973\,K $\Phi(q,t)$ can not consistently be described with 
the asymptotic scaling laws (Eq.\ \ref{betascal},\ref{tchq}). 
The dynamic range in which Eq.\ \ref{betascal} holds is increasing with decreasing
temperature.

\begin{figure}[t]
\includegraphics[scale=0.70]{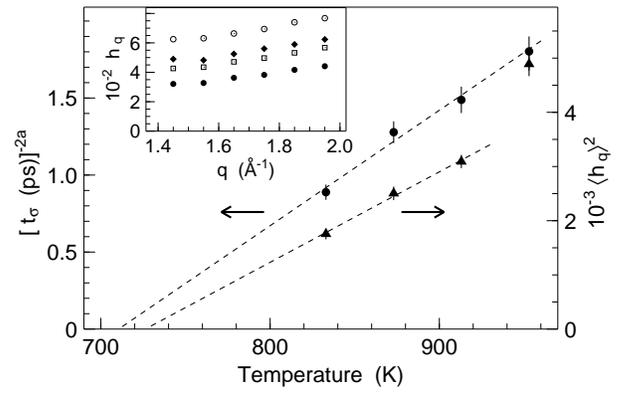}
\caption{\label{tcbeta} Mean amplitude $\langle h_q \rangle$ (triangles) and time scale $t_{\sigma}$
(circles) of fast $\beta$ relaxation in liquid Pd$_{43}$Ni$_{10}$\-Cu$_{27}$\-P$_{20}$ rectified
according the MCT predictions (Eq.\ \ref{tchq}), with an exponent $a=0.29$
defined by $\lambda=0.78$. 
The inset shows the $q$ dependence of the amplitude $h_q$ between 833\,K ($\bullet$) and
953\,K ($\circ$).
Both $h_q$ and $t_{\sigma}$ extrapolate to $T_c=720\!\pm\!20$\,K.
$\lambda$ and $T_c$ obtained from the analysis of the $\beta$ relaxation regime are
consistent with $T_c = 700\!\pm\!30$\,K and $\gamma=2.7\!\pm\!0.2$ from the analysis
of the mean $\alpha$ relaxation times (Eq.\ \ref{tauscal}). 
}
\end{figure}

The data were fitted in a two-step procedure:
starting with an arbitrary line shape parameter $\lambda$, 
fits to individual curves were used to estimate the scaling factors
$f_q$, $h_q$, and $t_\sigma$
which physically represent the Debye--Waller factor,
the amplitude of $\beta$-relaxation, 
and a characteristic time of $\beta$-relaxation.
Using these values, the $\Phi(q,t)$ measured at different temperatures
were superimposed on to master curves $(\Phi(q,t/t_\sigma)-f_q)/h_q$.
After fixing a $q$-independent mean $t_\sigma$,
the fit yielded a temperature and $q$ independent $\lambda\!=\!0.78\pm0.04$.

This result compares well to the hard spheres value $\lambda\!=\!0.766$
of the numerical MCT solution \cite{FuHL92},
to the $\lambda\!=\!0.77\pm0.04$ found in 
liquid Zr$_{46.8}$Ti$_{8.2}$\-Cu$_{7.5}$\-Ni$_{10}$Be$_{27.5}$ \cite{MeWP98}
and it is similar to the values in other viscous liquids \cite{Goe99,Tol01}.
$\lambda\!=\!0.78$ defines an exponent of the $\tau$ scaling law 
(Eq.\ \ref{tauscal}) of $\gamma\!=\!2.7$ \cite{Goe90}.
This is consistent with the temperature dependence of the mean relaxation times
that give $\gamma\!=\!2.7\pm0.2$ (Fig.\ \ref{diff}).

Figure \ref{tcbeta} shows amplitude $h_q$ and time scale $t_{\sigma}$
of the fast $\beta$ relaxation in liquid PdNiCuP rectified
according to Eq.\ \ref{tchq}.
The temperature dependence of $h_q$ and $t_{\sigma}$ is in accordance with 
the MCT predictions. Both $h_q$ and $t_{\sigma}$ extrapolate 
to $T_c=720\!\pm\!20$\,K that is close to the $T_c = 700\!\pm\!30$\,K
obtained via the $\tau$ scaling law (Eq.\ \ref{tauscal}).
The relaxational dynamics in liquid Pd$_{43}$Ni$_{10}$\-Cu$_{27}$\-P$_{20}$
can consistently be described with the universal scaling functions
of the mode coupling theory of the liquid to glass transition 
at temperatures up to $\simeq 1.3 \times T_c$ with
$T_c\!=\!710$\,K and $\lambda\!=\!0.78$.


\subsection{\label{matra}Atomic transport mechanism}

\begin{figure}[b]
\includegraphics[scale=0.78]{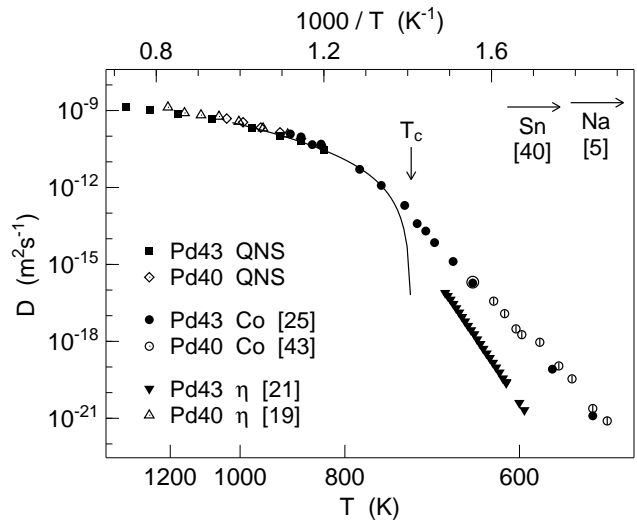}
\caption{\label{transall} 
Mass transport in PdNiCuP melts: At high temperatures diffusivities
are similar to diffusivities in liquid Sn and Na close to $T_m$ 
(marked by the arrows). 
Time scales for viscous flow and Co tracer diffusion start to decouple 
and Co tracer diffusivities merge into
an Arrhenius-type temperature dependence that extends down to 
the glass transition.
Both occurs in the vicinity of mode coupling $T_c$.
The line represents the MCT $\tau$ scaling law
with $T_c\!=\!710$\,K and $\gamma\!=\!2.7$.
}
\end{figure}

Pd$_{43}$Ni$_{10}$\-Cu$_{27}$\-P$_{20}$ exhibits an
excellent glass forming ability: a cooling rate of 
only 0.09\,K/s is sufficient to avoid crystallization
and to form a glass \cite{ScJB00}.
Consequently transport coefficients can continuously be measured 
from the equilibrium liquid down to the conventional glass
transition temperature $T_g$.
Figure \ref{transall} displays diffusivities in viscous PdNiCuP alloys
from Co tracer diffusion measurements,
that cover more than 13 decades \cite{Zoe02,ZoRF02}, 
derived from the mean relaxation times and from viscosity \cite{HaBD02,LuGF02}
calculated via Stokes-Einstein (Eq.\ \ref{stoke}).

At 1373\,K the 
diffusivity in liquid PdNiCuP
is $1.4\,\times\,10^{-9}$\,m$^2$s$^{-1}$
which is similar to diffusivities known from monoatomic liquid metals
at their respective melting temperature.
E.g.\ close to the melting point inelastic neutron scattering reveals 
$D=4.2\,\times\,10^{-9}$\,m$^2$s$^{-1}$ in liquid sodium \cite{MoGl86} 
and tracer diffusion under $\mu$g conditions
$D=2.0\,\times\,10^{-9}$\,m$^2$s$^{-1}$ in liquid tin \cite{FrKW87}.
At high temperatures the Stokes-Einstein relation holds in liquid PdNiCuP 
(Fig.\ \ref{diff}).
On lowering the temperature diffusion slows down drastically. 
Time scales for viscous flow and Co tracer diffusion start to decouple 
in the vicinity of the mode coupling $T_c\simeq710\,$K
and differ already by more than 3 orders of magnitude
at 600\,K.
Also around $T_c$ Co tracer diffusivities merge into
an Arrhenius-type temperature dependence 
($D(T) = D_0 \exp{(-H/k_BT)}$, where $D_0$ is a prefactor and
$H$ the activation enthalpy)
that extends down to 
the glass transition \cite{Zoe02,ZoRF02}.

The temperature dependence of the transport coefficients in
Figure \ref{transall} strongly supports the MCT prediction 
of a change in the atomic transport mechanism from viscous 
flow to glass like hopping at $T_c$. 
There are extensive tracer diffusion measurements in 
supercooled Zr$_{46.8}$Ti$_{8.2}$Cu$_{7.5}$Ni$_{10}$Be$_{27.5}$
at temperatures between the glass transition and 
about 200\,K below $T_c$ \cite{diffrev}.  
The diffusivities of various tracers exhibit a size dependence:
the smaller the atoms the faster they diffuse and the smaller 
the activation enthalpy $H$.
In viscous Zr$_{46.8}$Ti$_{8.2}$Cu$_{7.5}$Ni$_{10}$Be$_{27.5}$
the diffusivities of the various atoms approach each other with increasing
temperature. 
Because in this alloy crystallization prevents access to  
the temperature range around $T_c$, one can not conclude
from the present data whether the diffusivities of the various tracers 
merge at the MCT $T_c$.

The isotope effect 
$E=(D_a/D_b-1)/(\sqrt{m_b/m_a}-1)$ (diffusivity $D$ and mass
$m$ of two isotopes $a,b$)
is a measure of the degree of collectivity of the atomic transport.
For diffusion via single jumps in dense packed lattices
$E$ is generally in the order of unity \cite{Meh90}.  
For uncorrelated binary collisions one also expects $E \to 1$.
A vanishing isotope effect indicates a collective transport
mechanism involving a large number of atoms.
In liquid Sn an isotope effect of about 0.4 has been reported, that
is increasing with increasing temperature \cite{FrKW87}. 
MD simulations on a binary Lennard-Jones liquid demonstrate
that changes in the density in the order of 20\,\%
only result in a continuous increase in $E$ from 0 to $\simeq$\,0.3
\cite{Sch01}.
In metallic glasses \cite{FaHR90} as well as in supercooled 
Zr$_{46.8}$Ti$_{8.2}$Cu$_{7.5}$Ni$_{10}$Be$_{27.5}$ \cite{EhHR98}
and Pd$_{40}$Ni$_{10}$\-Cu$_{30}$\-P$_{20}$ \cite{ZoRF02} 
the isotope effect is close to zero
as a result of a highly collective, thermally activated hopping process.
Measurements of the isotope effect in Pd$_{43}$Ni$_{10}$\-Cu$_{27}$\-P$_{20}$
\cite{Zoe02} give $E\simeq0.05$ over the entire temperature
range from the glass transition to the equilibrium liquid.

The atomic transport mechanism in dense
PdNiCuP remains highly collective
even in the equilibrium liquid 
in contrast to the findings in simple metallic liquids.
Viscous PdNiCuP exhibits an expansion
coefficient of $4\times10^{-5}$\,K$^{-1}$ \cite{LuGF02}.
Between 1373\,K and 833\,K, the density and therefore, 
assuming a temperature independent hard sphere
radius, the packing fraction changes only by about 2\,\%. 
An increase in the radius of the attributed hard sphere radii \cite{Pas88} 
with increasing temperature results in an even smaller change in the
packing fraction.
This small increase in the packing fraction causes the drastical slowing down 
of dynamics on temperature decrease in accordance with the mode coupling scenario.
Dense packing and the resulting collective transport
mechanism extends also to high temperatures,
where diffusivities are similar to that in simple metallic liquids.
This indicates that the atomic transport mechanism in liquids is not 
controled by the value of the transport coefficients but rather
by that of the packing fraction.

\section{Conclusions}

Atomic transport in liquid Pd$_{43}$Ni$_{10}$\-Cu$_{27}$\-P$_{20}$ has been 
investigated with inelastic, incoherent neutron scattering.
The PdNiCuP melt is characterized by a packing fraction 
of about 0.52 that is some 20\,\% larger as compared to monoatomic alkali-melts.
The self correlation function shows a two-step decay as known from other
non-metallic glass-forming liquids: a fast relaxation process
precedes structural relaxation. 
The structural relaxation exhibits stretching with a stretching exponent $\beta\simeq0.75$
and a time-temperature superpostion that holds for temperatures as high as $1.75 \times T_c$. 
The relaxational dynamics can consistently be described
within the framework of the mode coupling theory of the liquid to glass transition
with a temperature and $q$ independent line shape parameter $\lambda\simeq0.78$. 
Universal MCT scaling laws extrapolate to a critical temperature 
$T_c\simeq710$\,K some 20\,\% below the liquidus.

Diffusivities derived from the mean relaxation times 
compare well with Co diffusivities from tracer diffusion measurements.
Above $T_c$ diffusivities calculated from viscosity \cite{HaBD02} via the 
Stokes-Einstein relation are in excellent agreement with 
the diffusivities measred by tracer diffusion \cite{Zoe02} and neutron scattering.
In contrast to simple metallic liquids the atomic transport in dense
liquid PdNiCuP
is characterized by a drastical slowing down of dynamics on approaching $T_c$,
a $q^{-2}$ dependence of the mean relaxation times at intermediate $q$ 
and a vanishing isotope effect as a result of a highly collective transport mechanism
in the dense packed liquid.
At temperatures as high as $2\!\times\!T_c$
diffusion in liquid PdNiCuP
is as fast as in simple monoatomic liquids at their melting points.
However, the difference in the underlying atomic transport mechanism 
indicates that the diffusion mechanism in liquids is not
controlled by the value of the diffusivity but rather by that of the
packing fraction.

\begin{acknowledgments}
It is a pleasure to thank Winfried Petry and Helmut Schober 
for their support, Wolfgang G\"otze, Walter Schirmacher
and Walter Kob for a critical reading of the manuscript, Joachim Wuttke
for his ingenious data analysis program,
the Institut Laue--Langevin for the beamtime on IN\,6, Stephan Roth
for his help during the measurement
and Helga Harlandt for her help with the sample preparation.
This work is funded by the German DFG (SPP Phasenumwandlungen
in mehrkomponentigen Schmelzen) under project Me1958/2-1.
\end{acknowledgments}

\end{document}